# MatPilot: an LLM-enabled AI Materials Scientist under the Framework of Human-Machine Collaboration

Ziqi Ni[1], Yahao Li[1], Kaijia Hu[1], Kunyuan Han[1], Ming Xu[2], Xingyu Chen[1], Fengqi Liu[1], Yicong Ye[1, *], Shuxin Bai[1]

[1]*College of Aerospace Science and Engineering, National University of Defense Technology, Changsha, 410073, China*

[2]*College of Intelligence Science and Technology, National University of Defense Technology, Changsha, 410073, China*

\* Corresponding author.
E-mail: niziqi18@163.com (Ziqi Ni), 18505993519@163.com (Yicong Ye).

## Abstract

The rapid evolution of artificial intelligence, particularly large language models, presents unprecedented opportunities for materials science research. We proposed and developed an AI materials scientist named MatPilot, which has shown encouraging abilities in the discovery of new materials. The core strength of MatPilot is its natural language interactive human-machine collaboration, which augments the research capabilities of human scientist teams through a multi-agent system. MatPilot integrates unique cognitive abilities, extensive accumulated experience, and ongoing curiosity of human-beings with the AI agents' capabilities of advanced abstraction, complex knowledge storage and high-dimensional information processing. It could generate scientific hypotheses and experimental schemes, and employ predictive models and optimization algorithms to drive an automated experimental platform for experiments. It turns out that our system demonstrates capabilities for efficient validation, continuous learning, and iterative optimization.

## Keywords
AI materials scientist; Large language models; autonomous experimentation platform

## 1 Human-machine collaboration framework

In recent years, artificial intelligence (AI) has driven a revolutionary transformation in materials science. Data-driven approaches have significantly advanced materials research, leading to notable improvements in predicting material properties[1–4], optimizing compositions and experimental conditions[5–10], as well as discovering new materials[11–15]. However, this data-driven paradigm has inherent limitations: it tends to follow a linear logic, constructing a semi-closed mechanical system that relies heavily on human input. Additionally, it faces significant challenges in developing interpretable models, enhancing data learning efficiency, and elucidating complex structure-property relationships in materials. More critically, current AI methods applied in materials tend to overemphasize correlations while neglecting causal relationships. In many instances, simple causal inference guided by domain expertise can provide deeper insights than analyzing massive datasets. Without incorporating common-sense reasoning and deep domain knowledge, AI systems that rely solely on statistical analysis struggle to reach the level of human intelligence needed to discover complex scientific theories.

The emergence of large language models (LLMs) such as ChatGPT has opened the door to human-machine collaboration through natural language interaction. By enabling communication in natural language, humans can interact with AI agents to exchange specialized knowledge in materials science. When researchers' intuition and experience are systematically integrated, AI models can continuously learn and think through feedback. Engaging in scientific research through human-machine collaboration, where "humans are in the loop", allows for combining the computational advantages of machines with the intuitive judgment capabilities of human beings, breaking through the limitations of purely data-driven methods. Fig. 1 depicts the human-machine collaboration framework implemented in our MatPilot, showcasing the collaboration between researchers and AI in a materials laboratory.

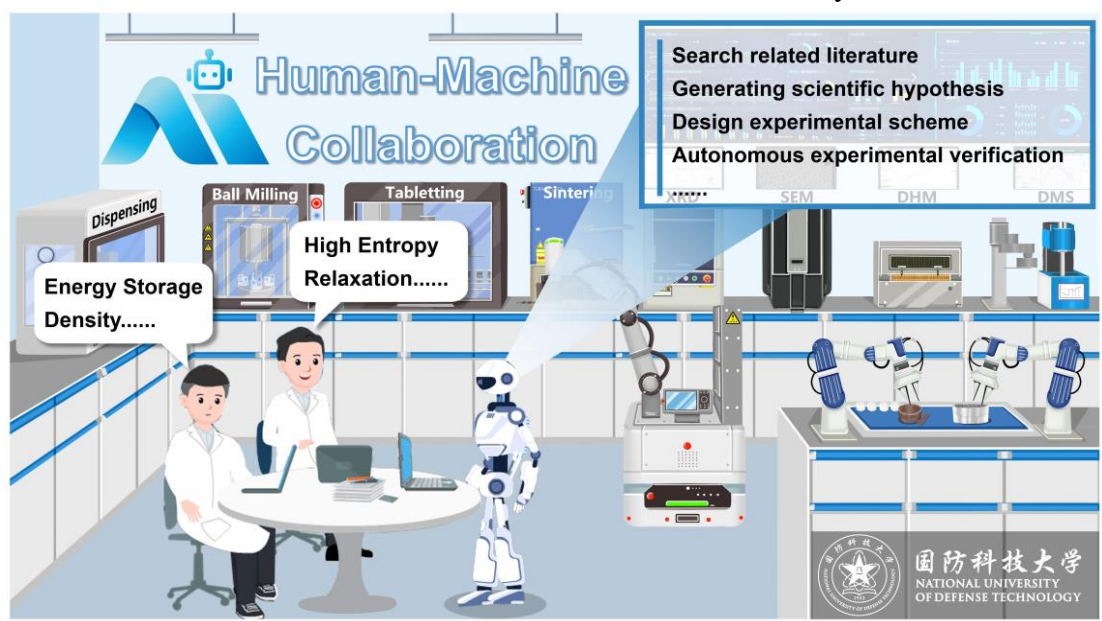

Fig.1 Human-machine collaboration framework implemented in our MatPilot.

## 2 MatPilot system architecture

Our group has developed MatPilot, an LLM-enabled AI materials scientist designed to enhance human-machine collaboration. It combines the strengths of human intuition with AI efficiency to enable a more systematic and effective exploration of materials science. MatPilot consists of a cognition module and a execution module (Fig.2), enhancing materials researchers' ability to think and perform efficiently. The cognition module, analogous to the human brain, is responsible for processing information, analyzing data, and making decisions. Meanwhile, the execution module resembles the body, tasked with performing the practical actions necessary for experimental procedures. Together, these two modules form a cohesive system where thinking and action are interlinked, enabling researchers to conceptualize, strategize, and implement their ideas in practice. This integration of cognition processing with physical execution establishes a research platform that supports both thinking capabilities and practical outcomes. By bridging the gap between abstract reasoning and tangible implementation, MatPilot facilitates a comprehensive approach to navigating the complexities of materials research.

MatPilot is designed as a tool to augment human creativity rather than replace it. The cognition and execution modules are intended to function autonomously while human researchers remain at the core of the scientific endeavor—their creativity, critical thinking, and intuition are irreplaceable drivers of innovation and discovery. MatPilot handles repetitive tasks, analyzes complex data patterns, and streamlines experimental workflows, allowing researchers to focus on more nuanced and conceptual work. Researchers can dedicate their time to hypothesis formation, interpretation of results, and refining research directions, with MatPilot deeply involved in and influencing the decision-making process. In this way, MatPilot acts as an intelligent collaborator, enhancing the researcher's ability for deep thinking and precise execution, while always keeping the human mind as the ultimate source of insight and direction.

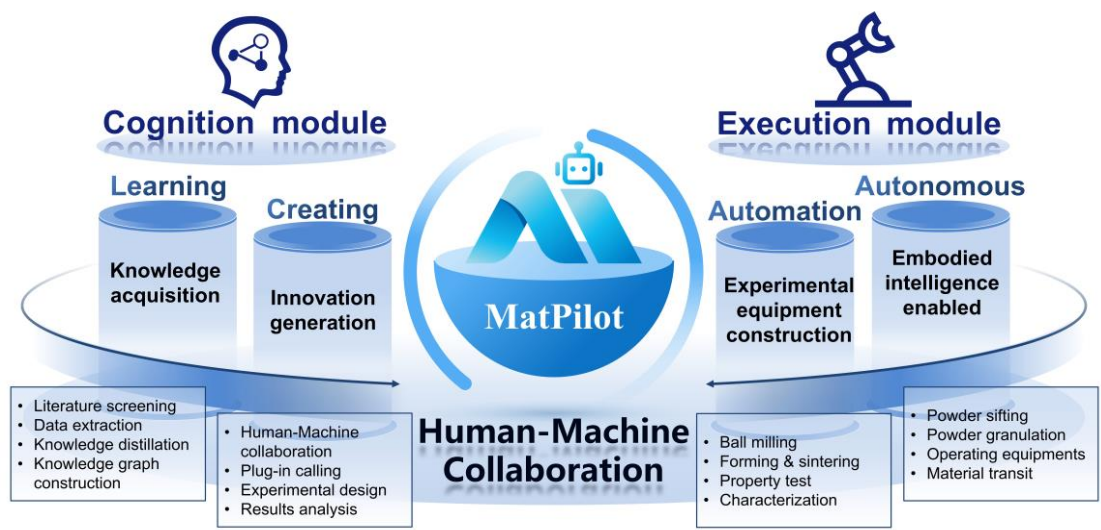

Fig.2 MatPilot's architecture (with energy storage ceramics as an example).

# 3 Cognition module

The cognition module of MatPilot integrates knowledge acquisition and innovation generation as its core functions. The knowledge acquisition function ensures that MatPilot continuously gathers the latest insights in materials science, while the innovation generation function enables it to propose novel research ideas and experimental designs. By combining these capabilities, MatPilot can efficiently understand existing research findings and assist in generating new insights, thereby providing substantial support to researchers.

## 3.1 Knowledge acquisition

We envision MatPilot evolving into an AI expert in materials science, with deep domain knowledge to effectively support researchers. To achieve this, we applied the retrieval-augmented generation approach, enabling a large language model to acquire specialized materials science knowledge through the retrieval of a high-quality knowledge base. The quality of this knowledge base is crucial for the effectiveness of the RAG method. Fig.3 illustrates the workflow for constructing such a high-quality knowledge base, which involves four main steps:

**a. Literature screening**

Initially, we filter a collection of core literature to identify the most relevant and core research findings, ensuring that the sources of information are of high quality and reliability.

**b. Data extraction**

Subsequently, we extract high-quality, structured data from the selected literature, focusing particularly on experimental procedures and key performance data. This step ensures that the extracted information is directly applicable to the model's reasoning and analysis tasks.

**c. Knowledge distillation**

Next, we condense complex scientific knowledge into manageable core concepts, enabling the model to efficiently grasp the essence of intricate problems and enhancing processing efficiency.

**d. Knowledge graph construction**

Finally, we construct a knowledge graph that elucidates the relationships between materials, processing methods, and performance attributes. This graph serves as a foundation for subsequent knowledge inference and relational analysis.

This approach offers two significant advantages. First, the system is no longer restricted to the static knowledge acquired during initial training; instead, it can continuously update its knowledge base to reflect the latest scientific advancements. Second, it can integrate multiple types of information (e.g., tabular data, distilled text, and relational graphs) and select the most suitable retrieval strategy depending on the specific nature of the query, thereby significantly enhancing retrieval efficiency and reasoning capabilities.

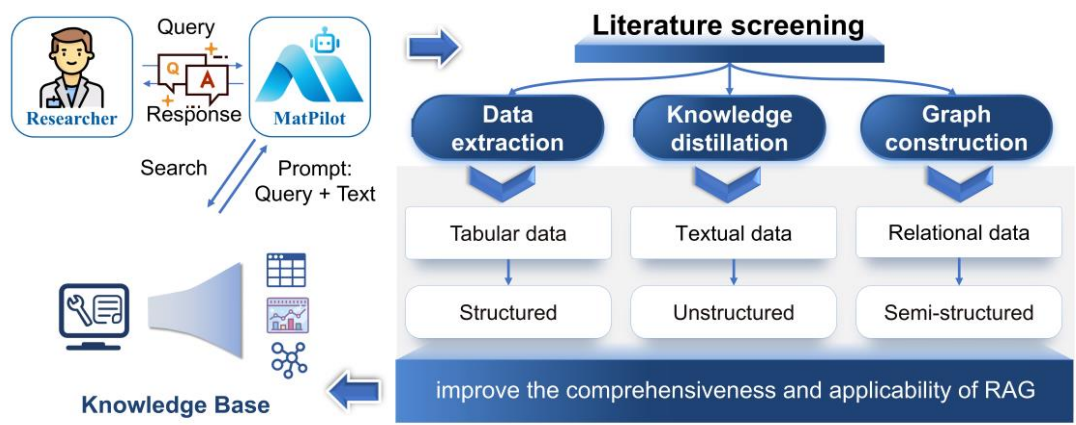

Fig.3 Workflow for constructing a high-quality knowledge base.

### 3.2 Innovation generation

MatPilot's creative capability is founded on the structural intelligence theory, which posits that innovation emerges from the synergy between divergent and convergent thinking. Divergent thinking enables the system to broadly explore the problem space and generate diverse solutions, while convergent thinking focuses on distilling these ideas into concrete, feasible proposals. Although LLMs excel at standard task processing, they often lack the imagination necessary for substantive innovation. To overcome this limitation, we have developed an innovation generation framework based on multi-agent and human-machine collaboration, as shown in Fig.4.

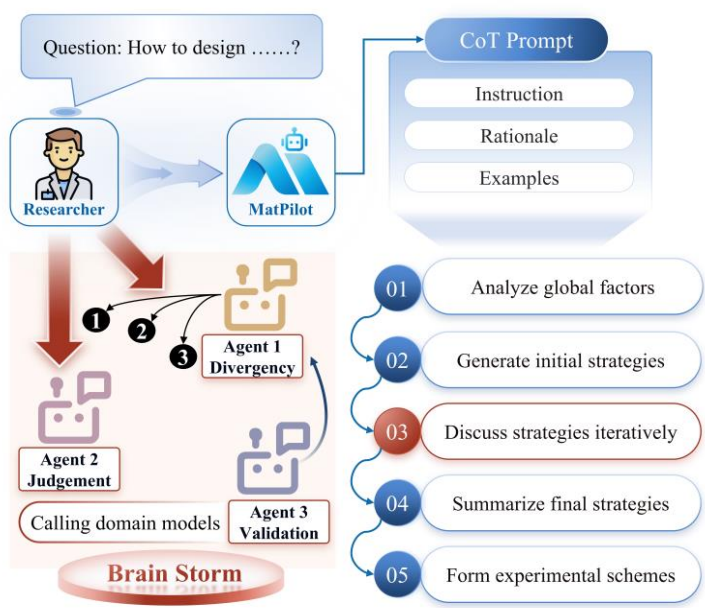

Fig.4 Multi-agent and human-machine debate collaboration framework for innovation generation.

In terms of multi-agent collaboration, the system incorporates three specialized types of agents: exploration agents, evaluation agents, and integration agents. Exploration agents are responsible for divergent thinking, generating diverse research directions through interdisciplinary knowledge association and heuristic reasoning. Evaluation agents focus on feasibility analysis, conducting comprehensive assessments across dimensions such as technical complexity, resource requirements, and expected outcomes. Integration agents coordinate perspectives among different agents, synthesizing disparate innovative elements into coherent research proposals. This multi-agent architecture emulates the collaboration of human research teams through continuous interactive dialogue and adaptive adjustments, significantly boosting the system's capacity for innovation.

The human-machine collaboration aspect establishes a bidirectional interactive mechanism. Human experts contribute domain knowledge, research experience, and strategic guidance, infusing the system's innovation process with high-level professional insights. In turn, the system harnesses its powerful data processing and analytical capabilities to rapidly generate multidimensional research directions for expert consideration. This process creates a positive feedback loop: expert feedback helps the system continuously optimize its innovation strategies, while the system's multifaceted analysis provides experts with novel research perspectives. This collaborative model ensures both the scientific validity and feasibility of innovative proposals.

MatPilot can not only generate innovative research directions but also design practical experimental protocols, thus playing a substantive creative role in materials science research. This framework enables the system to be an intelligent collaborative partner capable of research conceptualization and experimental design.

# 4 Execution module

In traditional materials research, researchers often need to perform numerous repetitive tasks, which are both labor-intensive and time-consuming. Experimental automation[16–20] has generated significant interest and has been applied by both academia and industry, particularly in fields such as pharmaceuticals and organic chemistry. In recent years, automation and autonomous experimental platforms in materials and chemistry have grown considerably. While there have been reports of autonomous experiments involving solid materials, a fully autonomous experimental platform that spans from preparation to characterization and performance testing has yet to be seen.

The execution module of MatPilot practically implements the research ideas and experimental plans developed by the cognition module. By leveraging automation and autonomous experiments platform, MatPilot's execution module effectively liberates researchers from these tedious experimental tasks, enabling them to devote more time to creative thinking and scientific inquiry. Furthermore, the execution module enhances experimental efficiency through the use of standardized and regulated procedures, thereby significantly improving the reliability and reproducibility of experimental results.

## 4.1 The entire automation process from material preparation to characterization

Materials research has traditionally been a resource-intensive endeavor, requiring both time and significant costs. The solid-state sintering method, widely used for ceramics, is a prime example of this complexity. The process involves numerous intricate steps, including raw material weighing, ball milling, sintering, granulation, property testing and so on. Each of these major steps consists of multiple sub-tasks, often requiring painstaking attention to detail. The conventional manual preparation of ceramic materials, therefore, becomes not only labor-intensive but also prone to variability and inconsistency.

MatPilot's execution module is transforming this by automating significant portions of the solid-state sintering workflow, as shown in Fig.5. By integrating automated workstations at every feasible point, the module reduces the need for manual intervention, ensuring consistency, precision, and accuracy across all critical stages. This strategic automation addresses key bottlenecks of the traditional process, especially maintaining consistency and efficiency in experiments.

The introduction of automation through MatPilot fundamentally reimagines the ceramic preparation process. Instead of a sequence of disconnected, laborious manual tasks, the workflow is transformed into an integrated, streamlined operation. This automation not only minimizes idle times between steps but also optimizes resource allocation, ensuring effective utilization of machinery, materials, and human oversight. The precision and consistency introduced by automation lead to highly reproducible and reliable experimental results, a crucial factor in materials research where even minor inconsistencies can lead to divergent outcomes.

By automating critical stages, MatPilot also enhances safety in the laboratory setting. Manual handling of ceramic powders and sintered materials often involves risks such as exposure to fine particulates or repetitive strain from laborious actions like ball milling. Automated systems mitigate these risks by taking over the most hazardous and repetitive parts of the process, allowing researchers to focus on higher-level tasks such as experimental design and analysis.

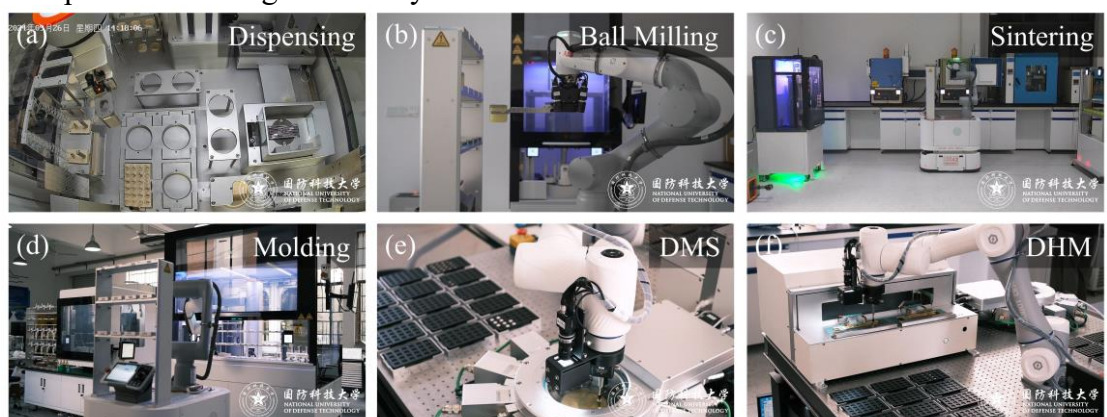

Fig.5 Automation workflow: (a) Dispensing; (b) Ball Milling; (C) Sintering; (d) Molding; (e) DMS; (f) DHM.

**4.2 Integration of embodied intelligence in autonomous experiment**

The laboratory environment was initially designed for human researchers, requiring substantial modifications to automate experimental processes. However, automating each piece of equipment to ensure compatibility with robotic arms would be both impractical and excessively costly. Moreover, traditional automation methods lack the precision, adaptability, and real-time feedback needed for complex tasks requiring high flexibility. This limitation is primarily due to the absence of necessary adaptive response capabilities and intelligent feedback mechanisms.

For instance, during the preparation process of ceramics, researchers rely on subjective judgment to evaluate particle quality and detect residuals in the sieving of powders. This reliance on human intuition poses challenges in adjusting the amplitude or frequency of robotic arm movements, which requires real-time perception and decision-making. Such subjective evaluations hinder the possibility of fully automating these processes, as conventional automation systems struggle to effectively handle these nuanced complexities.

Embodied intelligence emphasizes the ability of automation systems to adapt flexibly in uncertain environments through enhanced perception and interaction capabilities. And recent advancements, such as ALOHA[21] and ReKep[22], offer promising solutions to address our challenges. Mobile ALOHA performs a variety of complex household tasks autonomously or via remote operation, using imitation learning. ReKep leverages a vision-language model, GPT-4o, and relational keypoint constraints to enable robots to operate in complex environments, such as pouring tea.

The similarity between these household operations and our experimental tasks gives us great confidence in the possibility of achieving autonomous experiments in materials. We plan to implement embodied intelligence technologies over the next 1-2 years. Fig.6 shows some of the actions we are currently developing. This technology will enable us to design more intelligent systems that can dynamically adjust operations based on real-time feedback, thereby enhancing both the efficiency of experiments and the reliability of results.

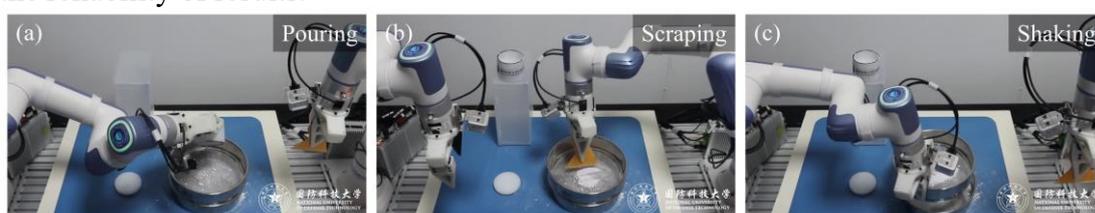

Fig.6 Actions empowered by embodied intelligence technologies for autonomous experiments: (a) pouring; (b) Scraping; (c) Shaking.

## 5 Discussion and outlook

MatPilot implements a highly efficient iterative optimization, wherein the cognition module and execution module collaborate continuously to generate hypotheses, conduct experiments, and integrate feedback. This process follows the concepts of evolutionary optimization: each iteration systematically refines experimental parameters based on empirical outcomes, achieveing in building on the cumulative knowledge accrued from prior iterations. Such iterative refinement facilitates the verification of scientific hypotheses and drives the ongoing enhancement of experimental strategies. By leveraging the insights from each experiment, MatPilot accelerates materials discovery by directing research efforts toward the most promising areas, optimizing resource allocation, and minimizing time spent on less rewarding experiments.

MatPilot serves as a materials research copilot, providing specialized support while ensuring that researchers maintain full control over their investigative journey. It means that each discovery is deeply interconnected to the human intellect behind it, making scientific exploration remaining an endeavor of curiosity, creativity, and profound understanding, which only human spirit can truly provide these qualities. The driving force behind scientific progress remains humanity's innate curiosity about the unknown and the relentless pursuit of truth. While MatPilot amplifies researchers' capabilities, it is ultimately the researchers who impart purpose and significance to every question, every experiment, and every breakthrough.

The rapid progression of AI is fundamentally reshaping the technological landscape, offering unprecedented opportunities in materials science. Motivated by this vision, we proposed and developed MatPilot, a platform that integrates AI into research workflows. We believe that human-machine collaboration will soon become an essential and natural part of scientific research. This will enable researchers to leverage AI for increasingly complex and demanding tasks, thereby maximizing efficiency, precision, and creativity in the research process. The synergy between human ingenuity and AI-powered exploration is creating unprecedented opportunities for innovation, meeting the growing needs of materials discovery and leading materials science into a new era.